\documentclass[conference]{IEEEtran}
\IEEEoverridecommandlockouts
\usepackage{cite}
\usepackage{amsmath,amssymb,amsfonts}
\usepackage{algorithmic}
\usepackage{graphicx}
\usepackage{textcomp}
\usepackage{xcolor}
\usepackage{siunitx}
\def\BibTeX{{\rm B\kern-.05em{\sc i\kern-.025em b}\kern-.08em
    T\kern-.1667em\lower.7ex\hbox{E}\kern-.125emX}}
\begin{document}

\title{BusTime: Which is the Right Prediction Model for My Bus Arrival Time?}

\author{
\IEEEauthorblockN{
Dairui Liu\IEEEauthorrefmark{1},
Jingxiang Sun\IEEEauthorrefmark{1}, and 
Shen Wang\IEEEauthorrefmark{2}}
\IEEEauthorblockA{
\textit{Beijing-Dublin International College, University College Dublin}, Dublin, Ireland.\IEEEauthorrefmark{1}\IEEEauthorrefmark{2}\\
\textit{School of Computer Science, University College Dublin}, Dublin, Ireland.\IEEEauthorrefmark{2}\\}
Email: \{dairui.liu, jingxiang.sun\}@ucdconnect.ie, shen.wang@ucd.ie
}

\maketitle

\begin{abstract}
With the rise of big data technologies, many smart transportation applications have been rapidly developed in recent years including bus arrival time predictions.
This type of applications help passengers to plan trips more efficiently without wasting unpredictable amount of waiting time at bus stops. 
Many studies focus on improving the prediction accuracy of various machine learning and statistical models, while much less work demonstrate their applicability of being deployed and used in realistic urban settings. 
This paper tries to fill this gap by proposing a general and practical evaluation framework for analysing various widely used prediction models (i.e. delay, k-nearest-neighbor, kernel regression, additive model, and recurrent neural network using long short term memory) for bus arrival time. 
In particular, this framework contains a raw bus GPS data pre-processing method that needs much less number of input data points while still maintain satisfactory prediction results. This pre-processing method enables various models to predict arrival time at bus stops only, by using a kd-tree based nearest point search method. 
Based on this framework, using raw bus GPS dataset in different scales from the city of Dublin, Ireland, we also present preliminary results for city managers by analysing the practical strengths and weaknesses in both training and predicting stages of commonly used prediction models.
\end{abstract}

\begin{IEEEkeywords}
bus arrival time prediction, computer performance, smart transportation, GPS data mining
\end{IEEEkeywords}

\section{Introduction}
\begin{figure}[htbp]
\begin{center}
\includegraphics[scale=0.35]{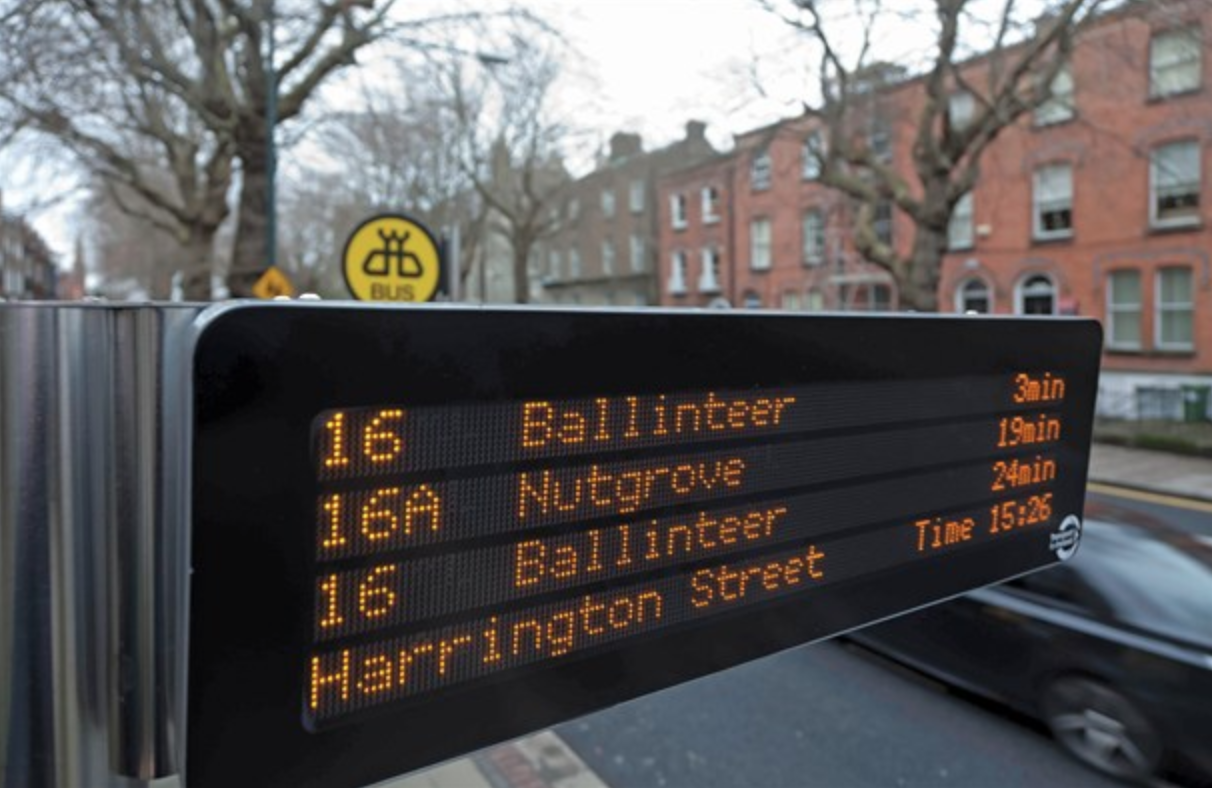}
\end{center}
\caption{Bus arriving time prediction system in Dublin, Ireland. (Source: itsinternational.com)}
\label{fig:dublin_bus}
\end{figure}

The big data era nowadays has seen an unprecedented amount of data generated than ever before, which enables smart decision making in all walks of life, including transportation \cite{nandan2013online}. 
For example, city managers in Dublin, Ireland, have collected large-scale GPS dataset of buses' trajectories for predicting their arriving time, as shown in Figure \ref{fig:dublin_bus}. 
GPS sensors are embedded on buses to report the real-time status (e.g. location and timestamp) periodically. 
Having collected these data, city managers can apply machine learning or statistical models to predict bus arrival times for potential passengers. 
Therefore, bus passengers can plan their trips more efficiently by avoiding excessive waiting time wasted at bus stops.

Many recent studies focus on improving the accuracy of the bus arrival time prediction. 
In 2011, k-nearest neighbors (k-NN) algorithm is applied to this problem \cite{Coffey2011}. 
This early research show that k-NN can provide predicted results with the mean absolute error less than about 3 minutes when the bus travelled at least around a quarter of its whole journey. 
A year later, the kernel regression (KR)\cite{Sinn2012} model is applied to improve k-NN by automatically selecting the proper number of most similar past trajectories for predictions, which have resulted in much stable arrival time estimations. 
Korm\'aksson et al. \cite{Kormaksson2014} proposed to use additive models, a more flexible tool that can combine multiple features and does not need GPS points interpolations to align different trips by fixed travelled distance. 
Last year, Pang et al. \cite{Pang2018} adopted recurrent neural network with long short-term memory block (RNN-LSTM) to capture long-term dependencies of bus arrival times at different location points, which further improves the prediction results.
Although the state-of-the-art bus arrival time predictions have been achieved more accurate and stable results, a performance analysis study using big data is missing for practitioners deploying them into realistic urban scenarios, by considering the trade-off between their accuracy and computation/storage requirement.

This paper tries to fill this gap by proposing a general and practical evaluation framework. In particular, this contribution can be divided into two parts as follows:
\begin{itemize}
    \item \textbf{An efficient bus GPS data pre-processing method that enables the comparison of many widely used prediction models.} To make the prediction results comparable, due to the fact that the raw GPS data is often noisy and have irregular update frequency, most of the existing models need to normalise the raw GPS dataset per bus trip by interpolating timestamps at every fixed travelled distance (e.g. every 100 or 50 meters)\cite{Sinn2012}. However, bus passengers normally do not care the prediction at the location which is not bus stop. The proposed data pre-processing method also interpolates arrival time, but at bus stops only, by using nearest-point-search. As the bus stops for a certain bus trip are at fixed locations, it is applicable for various existing prediction models. Normally, the number of bus stops are much less than the number of interpolated GPS points used in the literature, and our method using k-d tree \cite{kdtree1975} to accelerate the nearest point search. Therefore, it can reduce computation significantly and still maintained satisfactory prediction results.
    \item \textbf{A preliminary performance analysis of commonly used prediction models for practitioners.} Drawing on the aforementioned data pre-processing method, we evaluate five prediction models (delay, k-NN, KR, Additive, and RNN-LSTM) from the literature. To make this evaluation practical for city managers, we use a realistic dataset at different scales from the city of Dublin, Ireland, to demonstrate how their prediction accuracy and computation performance in training and predicting stages varies under different big data size. We summarize and analyze our evaluation results by comparing the mechanisms of these prediction models. We believe that these results can guide practitioners in deploying new smart prediction models in realistic urban settings.
\end{itemize}
\section{Data Pre-processing}

\subsection{Data overview}
We use open datasets which include bus schedules\footnote{https://data.gov.ie/dataset/gtfs-dublin-bus} and bus GPS\footnote{https://data.gov.ie/dataset/dublin-bus-gps-sample-data-from-dublin-city-council-insight-project} in the city of Dublin, Ireland. Bus schedule data, also known as General Transit Feed Specification (GTFS) data, specifies the route shapes, bus stops, and scheduled arrival times for each bus line. Bus GPS data contains about two months (i.e. November 2012 and January 2013) real-time status of buses such as timestamps (i.e. update frequency about 30 seconds), location, and relevant bus trip information. We pick bus line 46A (outbound, from Phoenix Park to Dun Laoghaire, also used in the literature \cite{Sinn2012}) to study as it is the most frequent bus line in our dataset and it is a long bus trip (i.e. nearly 20km) that gives us enough points to study.

We define a bus trip as a bus movement from its first stop to its last one serving its predefined schedule. We denote such a bus trip as $T$, and $T=\{g_0, g_1, ..., g_i, ..., g_{n-1}\}$, in which $g_0$ and $g_{n-1}$ stand for the origin and destination points of this bus, respectively; $g_i$ is a 2-tuple $(t_i, d_i)$ that denotes the $i$th GPS point which travels $t_i$ seconds and $d_i$ meters away from $g_0$.

\subsection{Data pre-processing}
Our proposed data pre-processing method contains two phases: \textit{trip segmentation} and \textit{points interpolation}.

\begin{figure}[htbp]
\begin{center}
\includegraphics[scale=0.45]{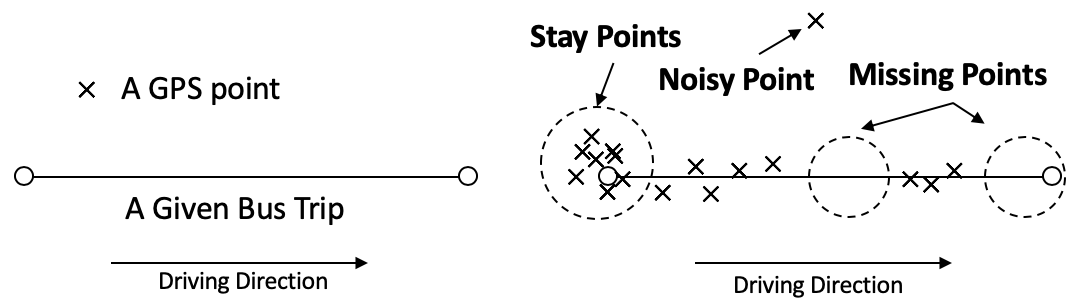}
\end{center}
\caption{Some well-known challenges in GPS data pre-processing.}
\label{fig:noisy_data}
\end{figure}

The main objective of trip segmentation is to identify each bus trip from raw GPS dataset with the knowledge of GTFS data (i.e. bus trip definition). There are some well-known challenges in GPS data pre-processing \cite{zheng2015trajectory} that our method needs to deal with. As shown in Figure \ref{fig:noisy_data}, \textit{stay points} means that there are a lot more densely distributed GPS points around a fixed location, when a bus is approaching, staying, or leaving its destination or origin stop. It becomes tricky to find exact point when the bus arrives, leaves or stays at such stops. Sometimes, especially in tunnels and high building areas where GPS signal is not stable, \textit{noisy points} can also be found which deviate spatially much further away from where they are supposed to be. Moreover, it is also common to have \textit{missing points} for a long distance or time duration due to sensor failures.

Inspired by the work in \cite{Wang2017}, we developed our trip segmentation method as follows:
\begin{enumerate}
    \item Coarse filter. There are two data fields in the raw GPS dataset: \textit{vehicle} and \textit{vehicle journey}, which combined means a specific bus vehicle that runs a certain trip. Although these data fields are normally not reliable, we can use them to coarsely narrow down the range of a bus trip to be filtered. Then, for each roughly split bus trip, we do the followings:
    \item De-duplication. There are many consecutive GPS points that have the exact location information but different timestamps. We remove such duplicated data points, which contributes to solve the \textit{stay points} issue.
    \item Further split by time. If the timestamp difference between two consecutive GPS points is higher than a predefined threshold (900 seconds in our case), we treat them as the last point of a trip and the first point of the next trip, respectively. This contributes to solve \textit{missing points} and \textit{noisy points} issues.
    \item Judge direction. We use the first 30 GPS points of a certain coarse bus trip to judge the direction of a trip. For the case of bus line 46A, the task is to judge if it is outbound or inbound, as they are the same route but different directions. Note this also implicates that we dump out a coarse bus trip if it has less than 30 points as it does not have enough points for further analysis. In particular, we use the first GPS point and the 30th GPS point as a vector to calculate the cosine similarity with two predefined bus trips from GTFS dataset. This step contributes to solve the \textit{stay points} and \textit{missing points} issues.
    \item Ensure completeness. We then further check if the first and last points of a certain rough trip is within the 300meters distance from its predefined origin and destination stops from GTFS dataset. We also ensure that the travel time and distance for the whole trip should not be less than the half of its predefined length and duration from GTFS dataset. This step overall contributes to avoid the \textit{missing points} issue.
\end{enumerate}

After getting a set of cleaned and segmented bus GPS trips, as normally done in the literature\cite{Sinn2012}, to make the prediction models work and comparable, we need to interpolate them, every 100 meters to align different GPS bus trips to a unified fixed travel distance, which we call it \textit{distance-based trip}.

\begin{figure}[htbp]
\begin{center}
\includegraphics[scale=0.40]{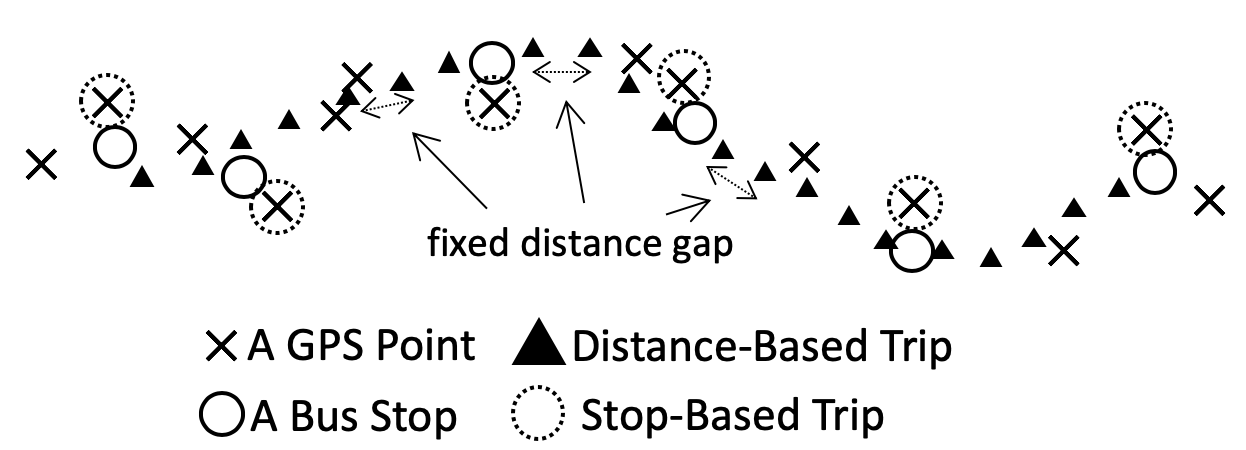}
\end{center}
\caption{The illustrative comparison between distance-based trip and stop-based trip.}
\label{fig:stop_based}
\end{figure}

As shown in the Figure \ref{fig:stop_based}, we propose a \textit{stop-based trip} interpolation method based on a practical view that bus passengers only care about arrival times at bus stops, rather than many interpolated points at locations with a fixed distance gap. Additionally, it slows down the training and prediction process as the number of interpolated points are often a lot more than the number of bus stops. Our proposed \textit{stop-based trip} interpolation method works as follows:

\begin{enumerate}
    \item Construct kd-tree for each bus trip. We use kd-tree to accelerate the nearest-point search process. We only use the location information (i.e. latitude and longitude) as the index to build kd-tree.
    \item Only select the GPS point that is the closest to each of its predefined bus stops. We loop through each bus stop sequentially in the predefined bus trip. For each bus stop iteration, we select its corresponding nearest GPS point in the GPS bus trip data. Then, to ensure all points in the resulted \textit{stop-based trip} are still in chronological order, all the previous data points of the selected nearest point in the GPS bus trip data are removed. Finally, the kd-tree is updated for the next iteration.
\end{enumerate}

For all trips, we list the main statistics of the distance deviation from the selected nearest GPS point to its corresponding bus stop. From Table \ref{tab:np_dist} we can see that this deviation is distributed around 33.53 meters on average with the standard deviation 111.72 meters. Moreover, there are 99\% points that only have less than 244 meters away from its nearest bus stop. If we take 30km/h as an average travel speed for a bus, then this 244 meters accounts for less than 30 seconds travel time error. Thus, we conclude that our interpolation introduces acceptable noise levels in practice in terms of deviated distance. We also want to highlight that this low deviated distance is due to our efficient and rigorous data pre-processing method, as it always gives us a complete bus trip data to interpolate.

\begin{table}[htbp]
\caption{The statistics of distance (meters) between the bus stop to its nearest bus GPS point (520,898 point pairs in total).}
\begin{center}
\begin{tabular}{|c|c|c|c|c|c|c|} \hline
 mean & std & 1\% & 25\% & 50\% & 75\% & 99\% \\ \hline
 33.53 & 111.72 & 0.31 & 2.82 & 6.45 & 36.95& 244.12 \\ \hline
\end{tabular}
\label{tab:np_dist}
\end{center}
\end{table}

\section{Evaluation Methodology}
We evaluate five prediction models: delay\cite{Sinn2012}, k-NN\cite{Coffey2011}, KR\cite{Sinn2012}, additive(BAM)\cite{Kormaksson2014}, and RNN-LSTM\cite{Pang2018}. We implemented them along with the whole evaluation framework in Python and make the code accessible at a Github repository\cite{git@2020}, in which you can also find our pre-processed testing data about bus line 46A outbound. In particular, due to the limitation of the additive model Python library pyGAM\footnote{https://pygam.readthedocs.io/en/latest/}, we only implement the basic additive model (BAM) to compare. Besides, because of the limited data accessibility, for RNN-LSTM, we only use the trip departure time, travel time, travel distance, and travel distance for the next point to construct a input vector. We implement RNN-LSTM using Tensorflow 2.0 and we use Adam optimiser with learning rate 0.01.

After pre-processing, we have selected 4311 trips that we can use for evaluation. The ``distance-based trip" interpolated from our testing dataset contains 191 points each trip, while our proposed ``stop-based trip" includes 59 points each trip only. In the evaluation, we will demonstrate that our ``stop-based trip" can reduce the computation time in both training and prediction process, while still maintained satisfactory accuracy.

In practice, city managers or engineers can only use historical data for training models to predict using real-time bus location updates, and these models will be re-trained after a certain time (normally, few weeks or months, as opposed to few seconds or minutes) using new data. Thus, we sort all 4311 trips in chronological order.

To assess the performance of these 5 models under various data scale, we divide all trips into 9 groups by choosing its first 500, 1000, 1500, 2000, 2500, 3000, 3500, 4000, and 4311 trips. Conventionally, for each of these 9 groups, we choose the first 80\% trips for training, and the rest 20\% trips for testing. We analyse the computation performance in training and prediction processes, respectively. Training process evaluation often indicates the amount of data that one can handle, while assessing prediction process tells us how well a model can be used in real-time settings.

Additionally, suppose we have $m$ trips and $n$ points for each trip, the predicted travelled time for the $i$th point in the $j$th trip is $\hat{t^j_i}$. To assess the model accuracy, the following metrics are used:

\begin{itemize}
    \item MAE: Mean absolute error.  This metric represents the average absolute deviation of the estimated travel time from the actual travel time in seconds, which is intuitive for people to understand. 
    \begin{equation} \label{eq:mae}
        MAE = \frac{\sum_{j}^{m}\sum_{i=1}^{n-1}|\hat{t^{j}_i}-t^{j}_i|}{m(n-1)}
    \end{equation}
    \item MAPE: Mean absolute percentage error. This metric represents MAE in percentage with respect to the realistic travel time. Normally, the absolute error tends to be large for predicting the travel time at the place which is far from the current location. However, as the actual travel time at such far distance from the origin location is also large, using MAPE can normalize this effect.
    \begin{equation} \label{eq:mape}
        MAPE = \frac{\sum_{j}^{m}\sum_{i=1}^{n-1}\frac{|\hat{t^{j}_i}-t^{j}_i|}{t^{j}_i}}{m(n-1)} \times 100\%
    \end{equation}
    \item RMSE: Root mean squared error. This metric penalizes the estimated travel time that deviates too much from its realistic one.
    \begin{equation} \label{eq:rmse}
        RMSE = \frac{\sum_{j}^{m}\sqrt{\sum_{i=1}^{n-1}{(\hat{t^{j}_i}-t^{j}_i)}^2}}{m(n-1)}
    \end{equation}
\end{itemize}

Note that to make the prediction process more realistic, we simulate this process as a real time one, in which when a bus just started its trip, it predicts arrival time at all the rest $n-1$ points / bus stops (the first point or bus stop is not required to be predicted as it is always 0). Generally, when a bus arrives the $i$th point / bus stop, it generates $n-1-i$ estimates for the rest points / bus stops. When a bus arrives the final point or bus stop, it does not need to estimate. Thus, for a certain bus trip, it generates $\frac{n^2}{2}$ travel time estimates. For simplicity, we did not reflect them in Equation \eqref{eq:mae}, \eqref{eq:rmse}, and \eqref{eq:mape}.

The testing machine has CPU 1.6 GHz Dual-Core Intel Core i5, memory 16 GB 2133 MHz LPDDR3, and SSD hard disk.
\section{Evaluation Results}
We present and analyse our evaluation results in this section. Specifically, for each of the following evaluation subsections, we firstly compare how 5 models perform, then demonstrate the superiority that our ``stop-based trip" interpolation method has over traditional ``distance-based trip". We present LSTM separately as it requires tuning many hyper-parameters manually. We only test LSTM on 1000 trips and 3000 trips.

\subsection{Training}
The model k-NN does not have a normal training process, as its key parameter $k$, the number of nearest trajectories, is normally tuned manually, as opposed to automatically trained by computer. As shown in Table \ref{tab:train_time}, the delay model requires the least amount of training time, which is 2 to 4 times faster than KR, and at least 10 thousand times faster than additive model and LSTM shown in Table \ref{tab:eva_lstm}. LSTM model requires the most computation during the training process. The time spent on tuning hyper-parameters is not included. Thus, in terms of training efficiency, we rank the 5 models as follows (left is the better, right is the worse):
\begin{center}
    Delay $>$ KR $>$ Additive $>$ k-NN $>$ LSTM
\end{center}

The results also show that with the increasing size of input data, our ``stop-based trip" can significantly reduce the amount of computation time required, up to about 5 times less, compared with the ``distance-based trip" method.

\begin{table}[htbp]
\caption{The training time of 3 bus arrival time prediction models using stop-based (left) and distance-based (right) trips. }
\begin{center}
\begin{tabular}{|p{0.5cm}|p{1.9cm}|p{1.9cm}|p{1.8cm}|} \hline
 &Delay(\num{1e-4}s) & KR(\num{1e-4}s)  & Additive(second) \\ \hline
500 & 2.17/2.76 & 4.09/8.95 & 1.35/4.96\\ \hline
1000 & 1.71/5.23 & 7.09/16.60 & 2.93/10.22\\ \hline
1500 & 2.89/7.37 & 10.00/24.77 & 4.58/15.71\\ \hline
2000 & 2.77/8.51 & 12.37/32.43 & 6.20/21.43\\ \hline
2500 & 3.89/11.90 & 13.79/36.99 & 7.82/27.40\\ \hline
3000 & 4.16/11.14 & 17.19/47.30 & 9.49/33.36\\ \hline
3500 & 4.44/15.48 & 19.98/58.15 & 11.12/39.43\\ \hline
4000 & 5.95/16.74 & 18.82/56.66 & 12.77/45.58\\ \hline
4311 & 6.05/16.78 & 24.70/76.88 & 13.92/49.34\\ \hline
\end{tabular}
\label{tab:train_time}
\end{center}
\end{table}

\begin{table}[htbp]
\caption{The evaluation results of RNN-LSTM using stop-based (left) and distance-based (right) trips. }
\begin{center}
\begin{tabular}{|p{1.4cm}|p{2.0cm}|p{2.0cm}|} \hline
 & 1000 (600 epoch) & 3000 (300 epoch) \\ \hline
Training (seconds)& 648.09/1969.79 & 950.12/3018.18 \\ \hline
Predicting (seconds)& 9.69/54.53 & 15.14/108.30 \\ \hline
MAE & 85.87/113.81 & 166.13/152.59 \\ \hline
RMSE & 128.47/164.63 & 194.36/180.10 \\ \hline
MAPE(\%) & 13.69/16.36 & 13.87/10.17 \\ \hline
\end{tabular}
\label{tab:eva_lstm}
\end{center}
\end{table}

\subsection{Predicting}
\begin{table}[htbp]
\caption{The prediction time (in seconds) of 4 bus arrival time prediction models using stop-based (left) and distance-based (right) trips.}
\begin{center}
\begin{tabular}{|p{0.5cm}|p{1.5cm}|p{1.5cm}|p{1.5cm}|p{1.5cm}|} \hline
 & Delay & k-NN & KR & Additive \\ \hline
500 & 0.56/0.18 & 0.34/1.06 & 1.57/8.11 & 0.07/0.21\\ \hline
1000 & 0.09/0.43 & 1.00/3.41 & 4.69/38.42 & 0.13/0.36\\ \hline
1500 & 0.14/0.80 & 2.04/6.86 & 10.30/92.02 & 0.19/0.53\\ \hline
2000 & 0.19/1.10 & 3.38/12.56 & 17.76/176.09 & 0.24/0.77\\ \hline
2500 & 0.25/1.44 & 5.12/16.52 & 32.3/275.53 & 0.29/0.90\\ \hline
3000 & 0.32/1.73 & 7.26/23.43 & 41.17/400.87 & 0.34/1.05\\ \hline
3500 & 0.36/2.21 & 9.89/30.86 & 63.16/576.68 & 0.38/1.21\\ \hline
4000 & 0.42/2.28 & 12.88/40.14 & 84.27/813.00 & 0.45/1.37\\ \hline
4311 & 0.49/2.90 & 14.88/53.77 & 86.49/996.42 & 0.50/1.57\\ \hline
\end{tabular}
\label{tab:predict_time}
\end{center}
\end{table}

As in the realistic city scenario, prediction demand occurs every second for many bus trips for the city servers where the prediction is deployed, the prediction service running time presented here is a aggregated one which have included all testing trips per testing group at all stops, rather than the average prediction time per trip per stop or point. The prediction time spent by delay model and additive model are the least among all compared models. Only a few seconds required at most makes these two models very practical to deploy. The prediction computation cost of KR increases sharply with the growing size of input data. LSTM can not achieve fast prediction partly due to the fact that we are not testing it on GPU/TPU machines\cite{jouppi2017tpu}, which are designed for accelerating the prediction process of deep learning models. Thus, in terms of predicting efficiency, we rank the 5 models as follows (left is the better, right is the worse):
\begin{center}
    Additive $>$ Delay $>$ k-NN $>$ LSTM $>$ KR
\end{center}

Under the full data setting 4311 trips, k-NN needs about 1 minute, which makes it less practical, but with our ``stop-based trip" method makes it more feasible to use by reducing the prediction time down to less than 15 seconds. KR is very sensitive to the big data. Traditional ``distance-based trip" method makes it almost infeasible to use (longer than 1.5minutes to predict) when the input data size is greater than 1500, while our method can still make it feasible to use under full data settings (less than 1.5minute to predict).

\subsection{Accuracy}
As shown in Table \ref{tab:mae}, \ref{tab:mape}, and \ref{tab:rmse}, in general, for all models, the more data they have, the better results they can achieve. However, the improvements are marginal for the most cases when the input size of data is more than 2000 trips. LSTM achieves the highest prediction accuracy which is obviously much better than others. But a larger dataset does not guarantee a higher accuracy for LSTM as it requires deep learning experts working on tuning many hyper-parameters for a given dataset. Although delay model has advantages in saving training and predicting computation cost, it does not show good prediction results. The accuracy of additive model prediction is the worst but it might because that we only implemented a simple version, and we use a new Python library rather than R used in literature. Besides, we can see that the accuracy clearly improves with the increasing amount of input data. Such a trend can not be directly seen in other models. Thus, in terms of predicting accuracy, we rank the 5 models as follows (left is the better, right is the worse):
\begin{center}
    LSTM $>$ KR $>$ k-NN $>$ Delay $>$ Additive
\end{center}

Our ``stop-based trip" method does not compromise a lot accuracy loss, while in some cases it can even improves accuracy (e.g. 1000 trips using LSTM).

\begin{table}[htbp]
\caption{The mean absolute error of 4 bus arrival time prediction models using stop-based (left) and distance-based (right) trips.}
\begin{center}
\begin{tabular}{|p{0.5cm}|p{1.5cm}|p{1.5cm}|p{1.5cm}|p{1.5cm}|} \hline
 & Delay & k-NN & KR & Additive\\ \hline
500 & 238.23/220.31 & 220.77/209.85 & 208.53/195.21 & 399.85/420.81\\ \hline
1000 & 237.75/224.07 & 227.10/217.11 & 216.29/201.10 & 306.11/313.32\\ \hline
1500 & 219.80/203.62 & 203.96/192.95 & 195.98/180.84 & 263.57/276.92\\ \hline
2000 & 203.70/193.13 & 198.92/191.16 & 187.47/177.00 & 261.22/262.47\\ \hline
2500 & 266.52/242.91 & 234.86/220.74 & 223.35/196.03 & 414.06/400.49\\ \hline
3000 & 224.55/209.11 & 211.29/197.94 & 196.70/179.48 & 300.00/301.83\\ \hline
3500 & 202.56/191.44 & 202.13/189.00 & 182.63/170.69 & 265.60/279.33\\ \hline
4000 & 208.43/196.25 & 198.83/190.90 & 186.46/173.24 & 275.80/278.72\\ \hline
4311 & 213.50/198.44 & 200.11/190.47 & 188.62/172.50 & 276.51/279.30\\ \hline
\end{tabular}
\label{tab:mae}
\end{center}
\end{table}

\begin{table}[htbp]
\caption{The root mean square error of 4 bus arrival time prediction models using stop-based (left) and distance-based (right) trips.}
\begin{center}
\begin{tabular}{|p{0.5cm}|p{1.5cm}|p{1.5cm}|p{1.5cm}|p{1.5cm}|} \hline
 & Delay & k-NN & KR & Additive \\ \hline
500 & 219.94/213.78 & 208.62/204.74 & 203.78/203.80 & 533.21/551.69\\ \hline
1000 & 219.32/215.13 & 206.92/207.46 & 200.5/202.91 & 434.99/431.44\\ \hline
1500 & 205.56/198.65 & 191.83/188.74 & 184.85/182.14 & 367.39/378.42\\ \hline
2000 & 198.51/196.34 & 191.69/192.85 & 181.70/182.76 & 348.96/353.03\\ \hline
2500 & 249.26/245.58 & 225.18/226.86 & 217.24/208.79 & 537.58/523.00\\ \hline
3000 & 217.49/214.73 & 206.40/204.76 & 193.46/190.35 & 399.22/402.80\\ \hline
3500 & 198.98/197.72 & 197.12/195.65 & 181.20/179.24 & 360.50/381.33\\ \hline
4000 & 203.38/198.77 & 194.39/193.95 & 182.77/179.37 & 372.68/380.03\\ \hline
4311 & 209.69/201.82 & 197.83/194.93 & 186.59/180.64 & 372.64/373.40\\ \hline
\end{tabular}
\label{tab:rmse}
\end{center}
\end{table}

\begin{table}[htbp]
\caption{The mean absolute percentage error (\%) of 4 bus arrival time prediction models using stop-based (left) and distance-based (right) trips.}
\begin{center}
\begin{tabular}{|p{0.5cm}|p{1.5cm}|p{1.5cm}|p{1.5cm}|p{1.5cm}|} \hline
 & Delay & k-NN & KR & Additive \\ \hline
500 & 10.69/9.78 & 9.82/9.10 & 9.29/8.49 & 29.81/32.74 \\ \hline
1000 & 8.70/7.76 & 8.21/7.46 & 7.92/6.90 & 30.37/22.95 \\ \hline
1500 & 8.14/7.12 & 7.44/6.65 & 7.24/6.27 & 21.93/19.16 \\ \hline
2000 & 9.33/7.00 & 8.58/6.89 & 8.05/6.29 & 51.22/18.39 \\ \hline
2500 & 12.21/10.29 & 10.77/9.28 & 10.17/8.21 & 36.83/25.64 \\ \hline
3000 & 9.53/8.28 & 8.64/7.64 & 8.26/6.97 & 25.83/20.78 \\ \hline
3500 & 8.30/7.59 & 7.88/7.21 & 7.43/6.62 & 25.86/22.25 \\ \hline
4000 & 8.25/7.46 & 7.86/7.14 & 7.35/6.51 & 29.34/21.36 \\ \hline
4311 & 8.68/7.54 & 8.03/7.12 & 7.63/6.47 & 29.85/20.15 \\ \hline
\end{tabular}
\label{tab:mape}
\end{center}
\end{table}

\subsection{Summary}
We give our review for all tested 5 models as follows:
\begin{itemize}
    \item Delay: A simple and efficient model to start with. It requires the least computation cost and still can give acceptable prediction results. This prediction accuracy can not be greatly improved by feeding more data.
    \item k-NN: It can generate better prediction results than delay model, but it requires hand tuning for parameter $k$. To achieve reasonable response time, when being deployed to predict in real time, the input data size should not be too much.
    \item KR: It can provide very good prediction results in all accuracy metrics and it can be trained very fast. However, when being deployed to predict in real time, to achieve reasonable response time, it needs to control the amount of input data size carefully.
    \item Additive: it is fast to be trained and deployed for prediction, as it only needs the trip departure time and current travel distance as the prediction inputs, rather than a partially traveled trajectory. If the data you have only contains travel distance and time, which means it is lack of multiple features (i.e. weather, hours of the day, etc), this model is not recommended due to low accuracy.
    \item RNN-LSTM: Very accurate prediction model. But it requires deep learning experts for tuning hyper-parameters and large amount of data to train. It is also better to be deployed at high-end machines designed for deep learning such as TPU/GPU to accelerate its training and predicting process.
\end{itemize}

Besides, we can summarize that our proposed ``stop-based trip" interpolation method outperforms conventional ``distance-based trip" method by significantly reducing the training and prediction computational cost, while still maintained satisfactory prediction accuracy.

\section{Conclusion and Future work}
To answer the question presented in the title: ``Which is the Right Prediction Model for My Bus Arrival Time?", this paper proposed a general and practical evaluation framework. Specifically, we proposed an efficient bus GPS data pre-processing method that enables assessing various models to predict arrival time at bus stops only. Besides, this method is used for assessing 5 existing prediction models using scalable big data at Dublin, Ireland. A performance analysis for practitioners is presented for these models, from the perspectives of training cost, prediction cost, and prediction accuracy.

We open sourced our evaluation framework and model implementations at Github \cite{git@2020}. In the future, we hope to work with people who are interested in improving this evaluation framework allowing assessing more bus routes from multiple cities with more prediction models to come.

% \section*{Acknowledgment}

% Authors would like to thank...

\bibliographystyle{IEEEtran}
\bibliography{IEEEabrv,main}

\end{document}